\documentclass[%
 reprint,
 superscriptaddress,
 amsmath,amssymb,
 aps,
prx,
onecolumn]{revtex4-2}

\usepackage{graphicx}
\usepackage{dcolumn}
\usepackage{bm}
\usepackage{amsmath} 
\usepackage{amsthm} 
\usepackage{amssymb}	
\usepackage{hyperref}
\usepackage{sidecap}
\usepackage{float}
\usepackage{comment}
\usepackage{mhchem}
\usepackage {color}
\usepackage{soul}
\usepackage[raggedright]{titlesec}
\titleformat{\section}
{\normalfont\Large\bfseries}{\thesection}{1em}{}
\titleformat{\subsection}
{\normalfont\large\bfseries}{\thesubsection}{0.6em}{}

\begin{document}

\preprint{APS/123-QED}

\title{Vortex confinement through an unquantized magnetic flux}

\author{Geunyong Kim}
\altaffiliation{Contributed equally to this work}
\affiliation{Department of Physics, Pohang University of Science and Technology, Pohang 37673, Republic of Korea}

\author{Jinyoung Yun}
\altaffiliation{Contributed equally to this work}
\affiliation{Department of Physics, Pohang University of Science and Technology, Pohang 37673, Republic of Korea}

\author{Jinho Yang}
\affiliation{Department of Physics, Pohang University of Science and Technology, Pohang 37673, Republic of Korea}

\author{Ilkyu Yang}
\affiliation{Department of Physics, Pohang University of Science and Technology, Pohang 37673, Republic of Korea}

\author{Dirk Wulferding}
\affiliation{Center for Correlated Electron Systems, Institute for Basic Science (IBS), and Department of Physics and Astronomy, Seoul National University (SNU), Seoul 08826, Republic of Korea}

\author{Roman Movshovich}
\affiliation{MPA-CMMS, Los Alamos National Laboratory, Los Alamos, New Mexico 87545, USA}

\author{Gil Young Cho}
\affiliation{Department of Physics, Pohang University of Science and Technology, Pohang 37673, Republic of Korea}

\author{Ki-Seok Kim}
\affiliation{Department of Physics, Pohang University of Science and Technology, Pohang 37673, Republic of Korea}

\author{Garam Hahn}
\affiliation{Pohang Accelerator Laboratory, Pohang University of Science and Technology, Pohang 37673, Republic of Korea}

\author{Jeehoon Kim}
\email{Corresponding author: jeehoon@postech.ac.kr}
\affiliation{Department of Physics, Pohang University of Science and Technology, Pohang 37673, Republic of Korea}




\date{\today}

\begin{abstract}

\end{abstract}

\maketitle
\noindent{\bf Abstract}\\
\nolinebreak\\
Geometrically confined superconductors often experience a breakdown in the quantization of magnetic flux owing to the incomplete screening of the supercurrent against the field penetration. In this study, we report that the confinement of a magnetic field occurs regardless of the dimensionality of the system, extending even to 1D linear potential systems. By utilizing a vector-field magnetic force microscope, we successfully create a vortex-antivortex pair connected by a 1D unquantized magnetic flux in ultra-thin superconducting films. Through an investigation of the manipulation and thermal behavior of the vortex pair, we uncover a long-range interaction mediated by the unquantized magnetic flux. These findings suggest a universal phenomenon of unquantized magnetic flux formation, independent of the geometry of the system. Our results present an experimental route for probing the impact of confinement on superconducting properties and order parameters in unconventional superconductors characterized by extremely low dimensionality. 

\section*{Introduction}

Superconductors are known to exclude magnetic fields from their interior, except for type \uppercase\expandafter{\romannumeral2} superconductors, which allow the entry of magnetic fields in the form of vortices. These vortices, referred to as fluxoids, consist of both magnetic fluxes induced by external magnetic fields and circulating supercurrents. Owing to the single-valued nature of the superconducting order parameter, the fluxoid must follow quantization. The quantized value of the fluxoid ($\Phi_{F}$) is determined by the sum of the magnetic flux ($\Phi_{B}$) and the supercurrent flux ($\Phi_{J}$), expressed as $\Phi_{F}=\Phi_{B}+\Phi_{J}=\Phi_{0}$, where $\Phi_{0}=h/2e$, with $h$ and $e$ are Plank's constant and the electron charge, respectively. When the contribution of the supercurrent  is negligible ($\Phi_{J}=$0), the quantization of the fluxoid reduced to the quantization of the magnetic field flux ($\Phi_{F}=\Phi_{B}=\Phi_{0}$). Conversely, in the presence of significant supercurrents, the quantized entity becomes the fluxoid rather than the magnetic field flux. When the magnetic flux is not quantized, the spatially-varying phase factor of the superconducting order parameter gives rise to various intriguing phenomena, including the Little-Park experiment in thin samples\cite{LittlePark1962}, transport current\cite{Harlingen1982,Ginzburg1989,Ginzburg1991}, Josephson effect\cite{Josephson1974,Likharev1979,Jaklevic1964}, as well as $p$-wave and $d$-wave pairing symmetry in unconventional superconductors\cite{Kirtley1999,Maeno2012,Kawabata2007}.       

Recently, there has been significant interest in studying the vortex confinement effect in low-dimensional superconductors as a means to investigate the inhomogeneous nature of superconducting order parameters\cite{Brandt1993,Stenuit2004,Engbarth2011,Geim2000,Zhang2022}. In confined two-dimensional (2D) disks, for instance, magnetic fluxes can form, but their flux values are smaller than $\Phi_{0}$ due to incomplete screening of the supercurrent against field penetration. Consequently, The quantization of magnetic flux is not satisfied in 2D confinement systems\cite{Geim2000,Zhang2022,Bardeen1961,Ginzburg1962,Fetter1980,Kogan1994}. Although unquantized magnetic fluxes give rise to a plethora of interesting superconducting phenomena, their investigation is challenging due to geometric limitations, making it difficult to create, visualize, and manipulate them. Specifically, the exploration of 1D confined fluxoids in conjunction with a purported 1D linear potential is particularly intriguing.     

In this article, we present the observation of 1D fluxoid confinement via an unquantized magnetic flux in ultrathin superconducting Nb films using vector-field, cryogenic magnetic force microscope (MFM). By utilizing the local stray magnetic field of an MFM tip, we generate a half-vortex ring, resulting in a vortex-antivortex dipole pair connected by a 1D confined fluxoid. Our study of the manipulation and thermal behavior of these confined vortex pairs through an unquantized magnetic flux in ultrathin films suggests that there are no dimensional restrictions on the formation of 1D superconducting fluxoids. 

\section*{Results}
\subsection*{Generation of a half superconducting vortex ring via an unquantized magnetic flux}
\begin{figure*}[t]
\centering
\includegraphics[width=8.8cm]{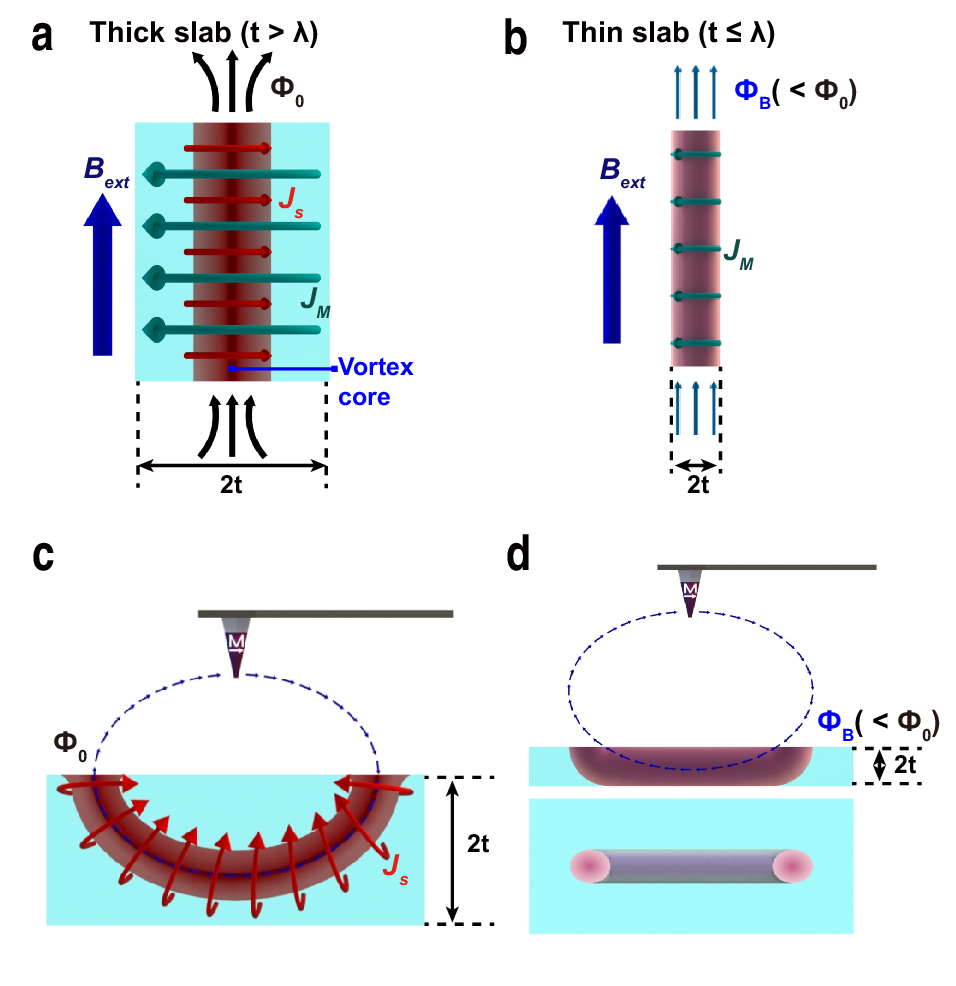}
\caption{Creation of a half vortex ring using a local dipole field. Fluxoid quantization of thick \textbf{a} and thin \textbf{b} slabs. The thick slab accommodates magnetic flux quantum ($\Phi_0$). The thin one experiences a field penetration but no vortex exists. Creation of a half vortex ring by a local dipole field of an MFM tip in thick \textbf{c} and thin \textbf{d} slabs. Unquantized magnetic flux can be trapped in the thin slab geometry.           
}
\label{fig1} 
\end{figure*}

In a thick slab of a superconductor under the influence of magnetic fields, the current and field distributions are depicted in Fig.~\ref{fig1}a. The green arrows represent the Meissner current ($J_M$) opposing the external field ($B_{ext}$), while the red arrows illustrate the shielding current ($J_s$) induced by the presence of superconducting vortices carrying a magnetic flux quantum ($\Phi_0$). On the other hand, in thinner geometries where the thickness is smaller than the magnetic penetration depth ($t < \lambda$), vortices are absent. However, in such systems, the magnetic flux remains confined within the film but is unquantized, characterized by a flux value of $\Phi_B$ that is less than the magnetic flux quantum $\Phi_0$, as depicted in Fig.~\ref{fig1}b. In a thick slab, it is possible to create a half vortex ring by employing a local magnetic dipole field, as shown in Fig.~\ref{fig1}c. It is noteworthy that this half vortex ring corresponds to a superconducting vortex-antivortex pair (SVAP) connected through a quantized magnetic flux ($\Phi_0$), with the associated shielding current ($J_s$). The line tension of the half vortex ring can be conceptualized as a one-dimensional spring exhibiting a linear potential. However, the situation changes in thin slabs. Consequently, the question arises as to whether SVAPs can still form with an unquantized magnetic flux ($\Phi_B$) and exhibit long-range interactions, as depicted in Fig.~\ref{fig1}d. Here, we demonstrate the formation of SVAPs with unquantized magnetic flux in ultrathin films, suggesting that superconducting vortices with arbitrary flux values can form regardless of the system's dimensions.

\subsection*{Creation of SVAPs using a local stray magnetic field of an MFM tip}
To investigate the properties of SVAPs, we employed a custom-built magnetic force microscope (MFM) equipped with a 2-2-9-T vector magnet, as previously described\cite{yang2016construction}. All experiments used a commercial MFM tip (PPP-MFMR from NANOSENSORS). The magnetic contrast in MFM images is expressed by the frequency shift $\Delta f$ of the MFM cantilever, which is directly linked to the force gradient $\partial F/\partial z$ with the relation  $\partial F/\partial z= -2k \Delta f/f_0$, where $k$ and $f_0$ are the cantilever’s spring constant and the bare resonance frequency of the MFM cantilever, respectively. The Nb film under study was deposited on a Si substrate using electron beam deposition and possesses the following characteristics: a thickness ($t$) of 300 nm, a superconducting transition temperature ($T_c$) of 8.8 K, a coherence length ($\xi$) of approximately 15.9 nm and a magnetic penetration depth ($\lambda$) of around 110 nm at 4.2 K, respectively. The values of $\xi$ and $\lambda$ were determined from the upper critical field ($H_{c2}$) (see Supplementary Note 1 for detailed methodology) and single-parameter simultaneous fits of the MFM signal\cite{nazaretski2009direct}, respectively. Thinner Nb films with thicknesses ranging from 30 to 100 nm were deposited on $\alpha$-Al${2}$O${3}$ substrates using a DC-magnetron sputtering system. These thinner films exhibited superconducting transition temperatures ($T_c$) ranging from 8.1 to 8.7 K. The magnetic penetration depths ($\lambda$) of these thinner films were estimated using the Meissner curve comparative method with MFM\cite{Kimlambda2012} and varied from approximately 145 nm to 90 nm (see Supplementary Note 1 for further details). 

\begin{figure*}[t]
\centering
\includegraphics[width=8.4cm]{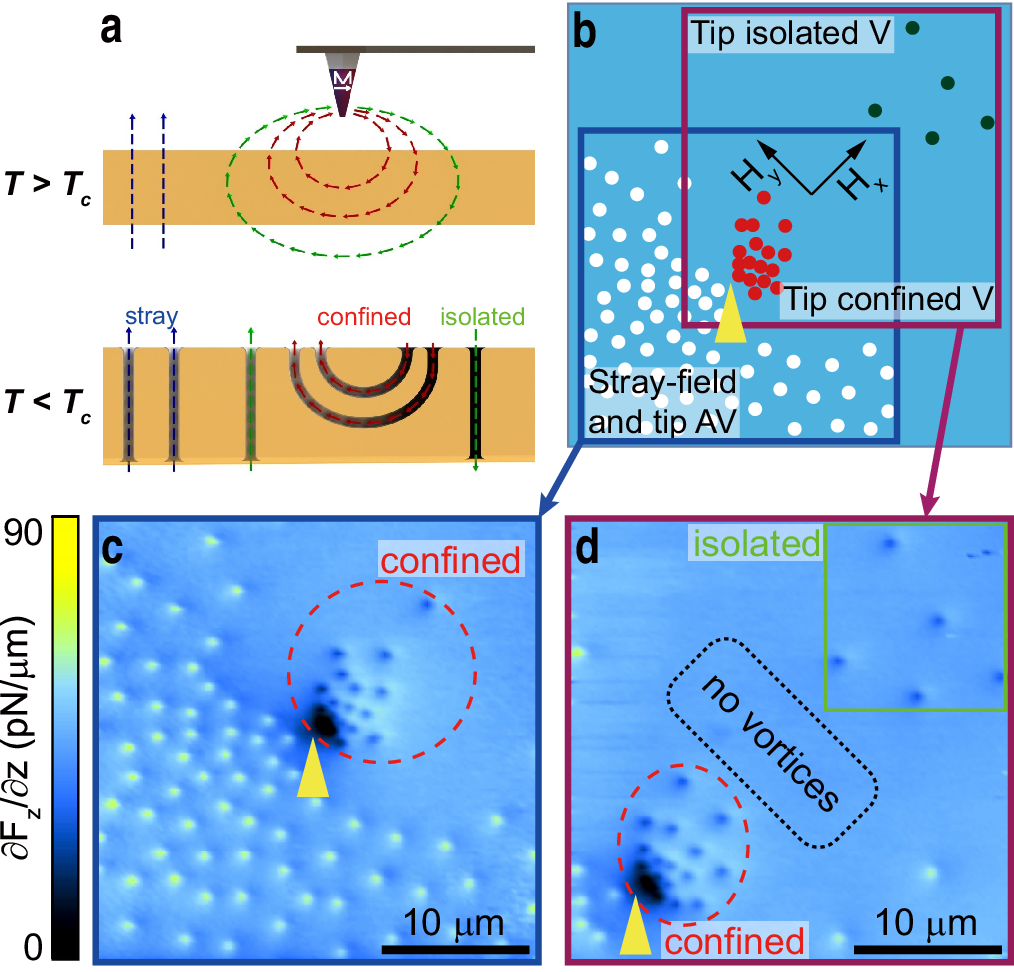}
\caption{Creation of a superconducting vortex-antivortex pair in MFM. \textbf{a} Schematics illustrating the creation of superconducting vortex-antivortex pairs (SVAPs) using an in-plane magnetized MFM tip. SVAPs (red dots) and trapped interstitial vortices (TIVs, green dots) are generated by the tip, while regular vortices (blue dots) are formed by the stray field of the system. \textbf{b} Illustration of the SVAP configuration created by the MFM tip. Antivortices (white dots) are induced by both the external stray field and the tip-field, while SVAPs (red dots) and TIVs (dark green dots) represent the confined vortices. The yellow arrow indicates the position of the MFM tip during the SVAP creation process. \textbf{c}, \textbf{d} MFM images obtained from two separate experiments with the same condition and shifted scan frame. The yellow arrow indicates the position of the MFM tip, which remains at a fixed tip-sample distance of 600 nm during the SVAP formation. Bright dots in the MFM image correspond to antivortices, while dark dots represent vortices.}
\label{fig2}
\end{figure*}

The presence of a local magnetic field induced by an MFM probe tip in close proximity to a superconducting film can give rise to the formation of a half fluxoid ring, as illustrated schematically in Fig.~\ref{fig2}a. This half fluxoid ring corresponds to a superconducting vortex-antivortex pair (SVAP) connected by a magnetic flux. It should be noted that in the absence of pinning, the SVAPs can annihilate due to the attractive line tension between the two ends. However, in real samples, the presence of a pinning landscape prevents their annihilation. The procedure for generating an SVAP using an MFM tip is as follows: (1) Above the critical temperature ($T_c$), the in-plane magnetized tip is positioned above the Nb film, generating a dipole magnetic field. (2) The sample is then cooled below $T_c$ to enter the superconducting state (see Supplementary Movie 1). Upon cooling, some of the U-shaped magnetic flux lines (red lines) originating from the tip become trapped within the superconducting medium, leading to the formation of SVAPs. The remaining flux lines (green lines) that penetrate through the backside of the sample give rise to tip-induced isolated vortices (TIVs) with a cylindrical shape. In addition to these tip-induced vortices, there is an uncompensated weak out-of-plane magnetic field (blue lines) arising from the stray field of our MFM system, which leads to the presence of isolated antivortices. Note that the principal origin of the system's stray field stems from the magnetization process of the MFM tip. Before we commence the MFM experiments at low temperatures, the tip undergoes a magnetization procedure using a superconducting magnet, subjecting it to a magnetic field of approximately 1 Tesla to enhance signal-to-noise ratio. As a consequence of magnetizing the MFM tip, a stray magnetic field permeates the entire MFM system, typically manifesting as a few gauss. This stray field influences the number of superconducting vortices under investigation.

An in-plane magnetized tip, polarized in the $+H_{x}$ direction by the superconducting vector magnet, was utilized to create SVAPs and TIVs, as depicted schematically in Fig.~\ref{fig2}b. Due to the limited scan size of our microscope scanner, we combined two scans to cover the region of interest, as shown in Fig.~\ref{fig2}b. The corresponding experimental data for the blue box and the red box in Fig.~\ref{fig2}b are presented in Figs.~\ref{fig2}c and ~\ref{fig2}d, respectively. Both images were obtained at 4.2 K using an in-plane magnetized tip\cite{yang2016construction}. Initially, our focus is on the vortices (dark dots) since distinguishing between the antivortices (bright dots) induced by the tip field (red and green lines) and the stray field (blue lines) is challenging at this stage due to their identical magnetic polarities. However, later on, the SVAP antivortices will be identified and distinguished based on their thermal behavior. As the SVAP vortex and antivortex are confined, the interaction force between them is expected to differ from that between the TIVs.

The first experimental evidence of a qualitative distinction between SVAPs and TIVs is the observation of a significant \textit{spatial gap} (indicated by the black dotted box) between the SVAPs (enclosed by the red circle) and the TIVs (enclosed by the green box), as illustrated in Figs.~\ref{fig2}d. In the normal state, prior to the medium transitioning into the superconducting phase, the spatial density of magnetic flux lines decreases gradually away from the MFM tip location. Note that the inhomogeneous spatial density of superconducting vortices is attributed to the superposition of the inhomogeneous magnetic field from the MFM tip and the system's homogeneous stray field due to the differential interaction between the SVAP and the stray-field-induced vortices. Upon the sample's superconducting transition, the ends of the SVAPs start moving towards their corresponding counterparts until they encounter strong pinning centers that counterbalance the attractive force from the other end of the SVAP. In contrast, the TIVs remain in their initial positions. Consequently, a spatial gap emerges, indicating the mobility of SVAPs and the existence of a long-range mutual attraction between the constituent ends.

\subsection*{The presence of the long range interaction of the SVAP}

\begin{figure*}[t]
\centering
\includegraphics[width=13cm]{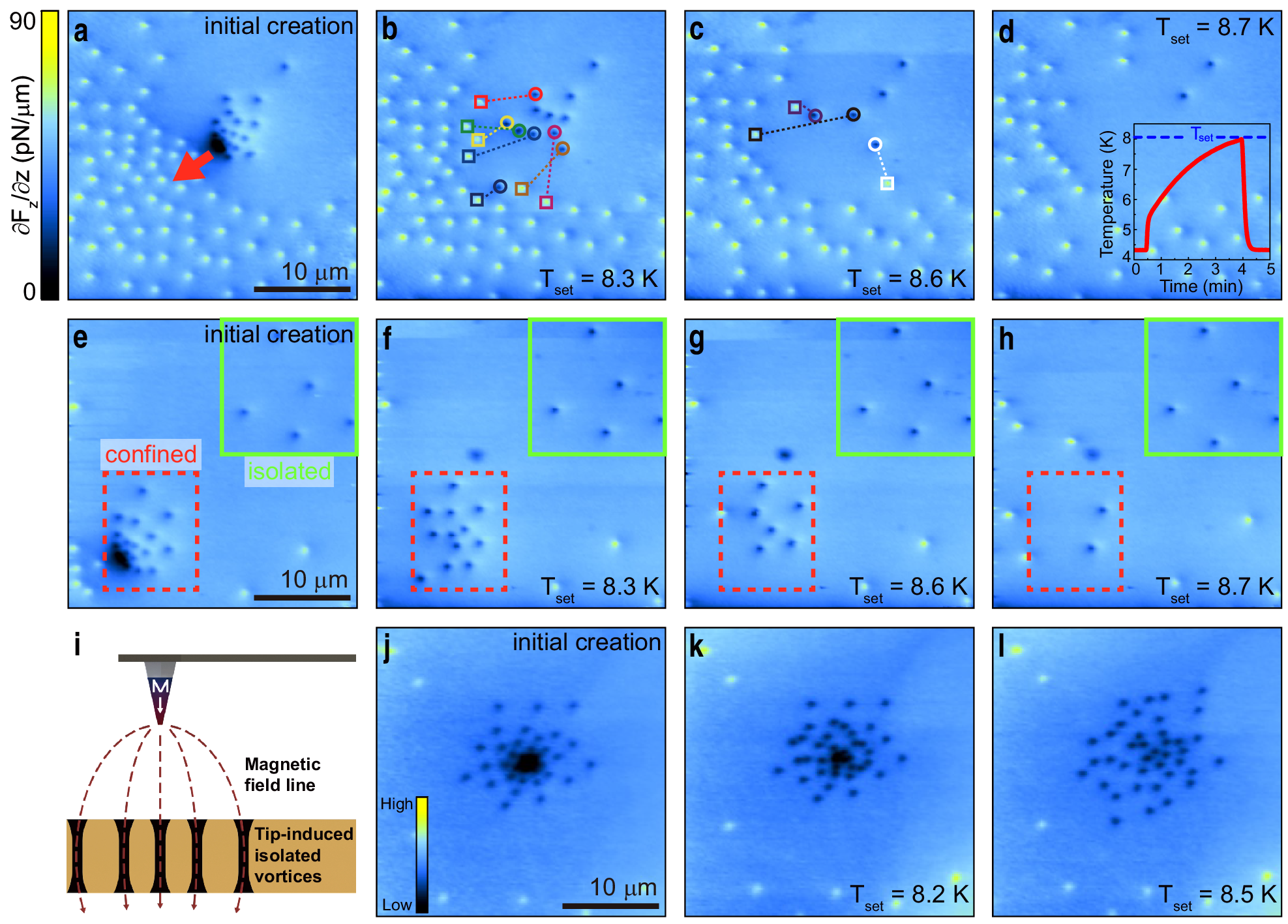}
\caption{Thermally assisted depinning of SVAPs. Thermal evolution of the vortices in Figs.~\ref{fig1}c and ~\ref{fig1}d are shown in \textbf{a}-\textbf{d} and \textbf{e}-\textbf{h}, respectively. After a thermal pulse with an amplitude of $T_{set}$, as denoted in the image, is applied, MFM images are obtained at 4.2 K. Inset curve in \textbf{d} provides a detailed time scale of the applied thermal pulse.The annihilation of each vortex-antivortex pair is traced by the box-circle link with a dotted color line. \textbf{i} A schematic view of creation of isolated vortices using an out-of-plane magnetization. \textbf{j}-\textbf{l} Thermal evolution of isolated vortices created according to the method described in \textbf{i}. The images demonstrate the isotropic behavior of the vortices during the thermal process. All MFM images were taken at 4.2 K and with a tip-sample distance of 600 nm.
}
\label{fig3} 
\end{figure*}

 To examine the nature of the interaction between the ends of the SVAPs, we applied heat pulses to the sample, allowing the SVAPs to overcome local pinning potentials\cite{Blatter1994}. As the thermal energy increased, one end of the SVAP approached the other end and settled at a stronger pinning potential, reducing the distance between them. This process was observed through sequential images in Figs.~\ref{fig3}a-~\ref{fig3}d (corresponding to the scan region in Fig.~\ref{fig2}c). The gradual shrinking and eventual annihilation of the SVAPs indicated the presence of an attractive interaction between the ends. Remarkably, this interaction occurred over a scale of several micrometers, significantly larger than the magnetic penetration depth ($\lambda \approx$ 110 nm) of the Nb film. The conventional isolated vortex-vortex interaction range of $\lambda$ alone cannot account for such long-range interaction\cite{tinkham2004introduction}. Therefore, the confinement of the SVAPs must be responsible for this observed long-range interaction.

Additional evidence for the confinement of the SVAPs was obtained through a comparative experiment between the TIVs and the SVAPs in Figs.~\ref{fig3}e-~\ref{fig3}h (corresponding to the scan region in Fig.~\ref{fig2}d). Unlike the SVAPs, the TIVs (green box) remained static even after multiple consecutive heat pulses close to the critical temperature ($T_c$), indicating the absence of a long-range interaction in the TIVs. The distinct thermal behavior of the SVAPs and TIVs further supports the presence of a long-range attractive interaction specifically between the ends of an SVAP.

To gain further insights into the interaction potential of an SVAP, we investigated the thermal behavior of isolated vortices created using an out-of-plane magnetized tip generated by a vector magnet, as shown in Figs.~\ref{fig3}i-~\ref{fig3}l. Initially densely packed, these isolated vortices gradually spread apart in a spatially isotropic manner with increasing heat pulses, as depicted in Figs.~\ref{fig3}j-~\ref{fig3}l. This behavior starkly contrasts with the confined vortices of the SVAPs, highlighting the distinct nature of the interaction between the ends of the SVAPs (see Supplementary movie 2 for further details).

\subsection*{Direct evidence of long range interaction in a single SVAP manipulation}
\begin{figure*}[t]
\centering
\includegraphics[width=13cm]{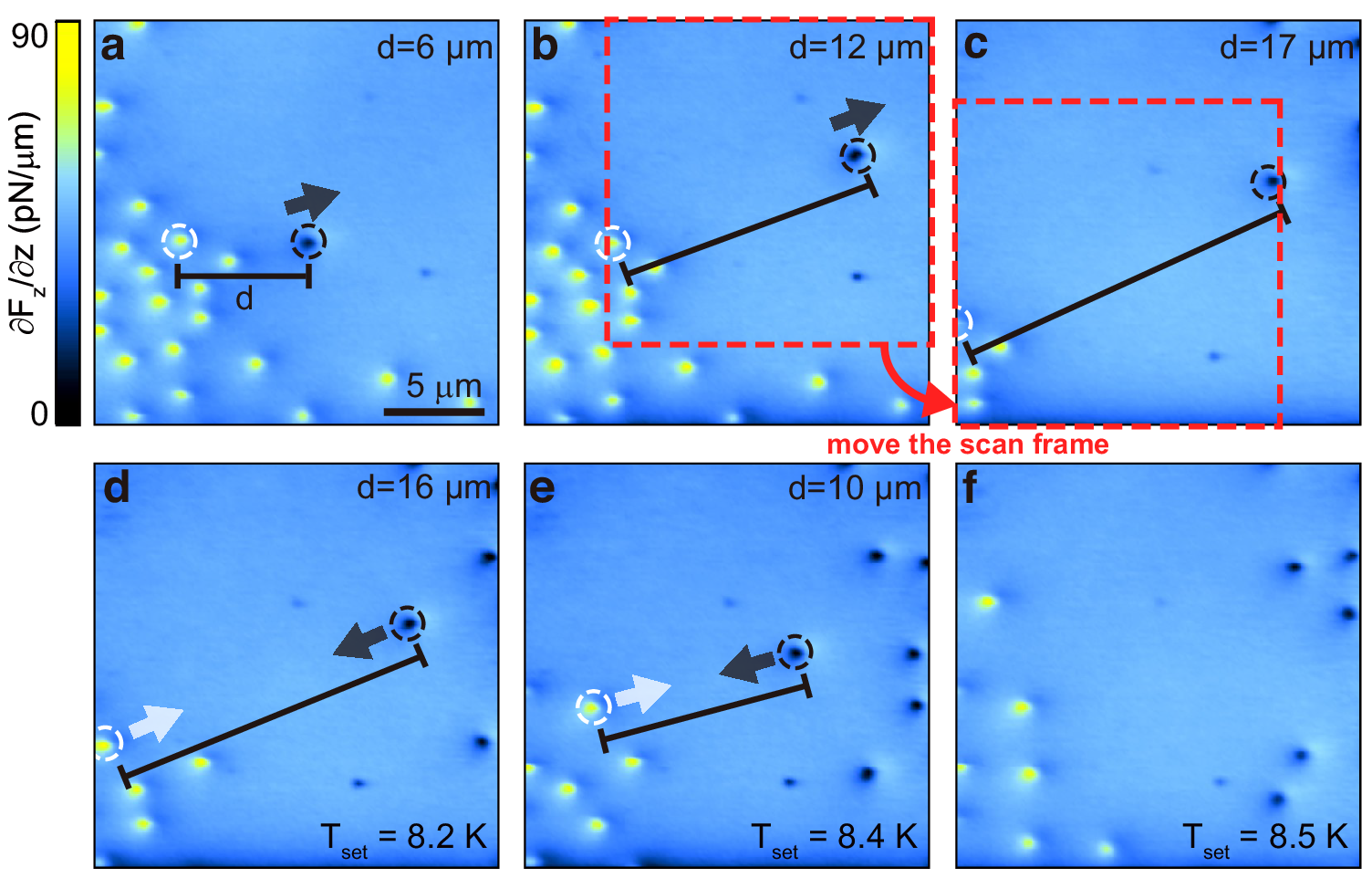}
\caption{Pair annihilation of an elongated single SVAP. \textbf{a}-\textbf{c} The manipulation process of one end of an SVAP by an MFM tip. After moving the scan frame step by step via piezo walkers, a final vortex-antivortex distance of 17 $\mu$m is achieved. \textbf{d}-\textbf{f} Thermally assisted depinning of both vortex and antivortex, and pair annihilation as a result of the successive heat pulses. Note that a confining interaction still exists between the antivortex and its previously manipulated counterpart because of the move of the antivortex as indicated with the white arrow, which is clearly seen in \textbf{d} and \textbf{e}. All MFM images were taken at 4.2 K and with tip-sample distance of 600 nm.
}
\label{fig4}
\end{figure*}

To directly investigate the presence of a long range potential, we present the manipulation of a {\it single} SVAP by an MFM tip and show its pair annihilation via thermal annealing (heat pulses). The starting point is prepared leaving behind only one end of an SVAP after moving the rest of the SVAPs by the tip stray field out of the scan frame, shown in Fig.~\ref{fig4}a. It is then continuously manipulated up to 17 $\mu$m, as shown in Figs.~\ref{fig4}b and ~\ref{fig4}c. The manipulation process has been carried out by scanning the single vortex with the tip-sample distance of 150 nm to enhance the interaction between vortex and tip\cite{Auslaender2009,Kremen2016,Ge2016,Wulferding2020}. To effectively manipulate a single vortex, the manipulation has been done at 7 K for decreasing the local pinning force, and then the result of the manipulation is verified by imaging the single vortex at 4 K. After the single vortex manipulation, we apply the same thermal annealing treatment with heat pulses as one described previously. The same gradual shrinking of the SVAP and its eventual annihilation are observed, as shown in Figs.~\ref{fig4}d-~\ref{fig4}f, clearly demonstrating that the SVAP remains intact through the entire MFM manipulation and thermal treatment, all the way until its annihilation.  The fact that the counterpart (marked as the white circle) moves toward the manipulated vortex (marked as the black circle) along the manipulated direction, as shown in Figs.~\ref{fig4}d and ~\ref{fig4}e, is a clear manifestation of a mutually attractive force.

The large distance over which the interaction persists, reaching up to 17 microns in our experiments, is particularly notable considering that it is over fifty times greater than the thickness of the Nb sample. This remarkable resiliency underscores the unique and intriguing nature of the SVAP and its confinement potential, which enables long range interations that cannot be explained by conventional vortex-antivortex interactions within superconducting systems. These findings contribute to our understanding of the fundamental properties of SVAPs and shed light on the underlying physics of confined superconducting vortex structures. The ability to manipulate and control individual SVAPs opens up new possibilities for exploring their behavior and harnessing their unique properties in future applications such as in the design of novel superconducting quantum technologies.

\subsection*{The confinement of the SVAP through an unquantized magnetic flux in thin films}
 
\begin{figure*}[t]
\centering
\includegraphics[width=13cm]{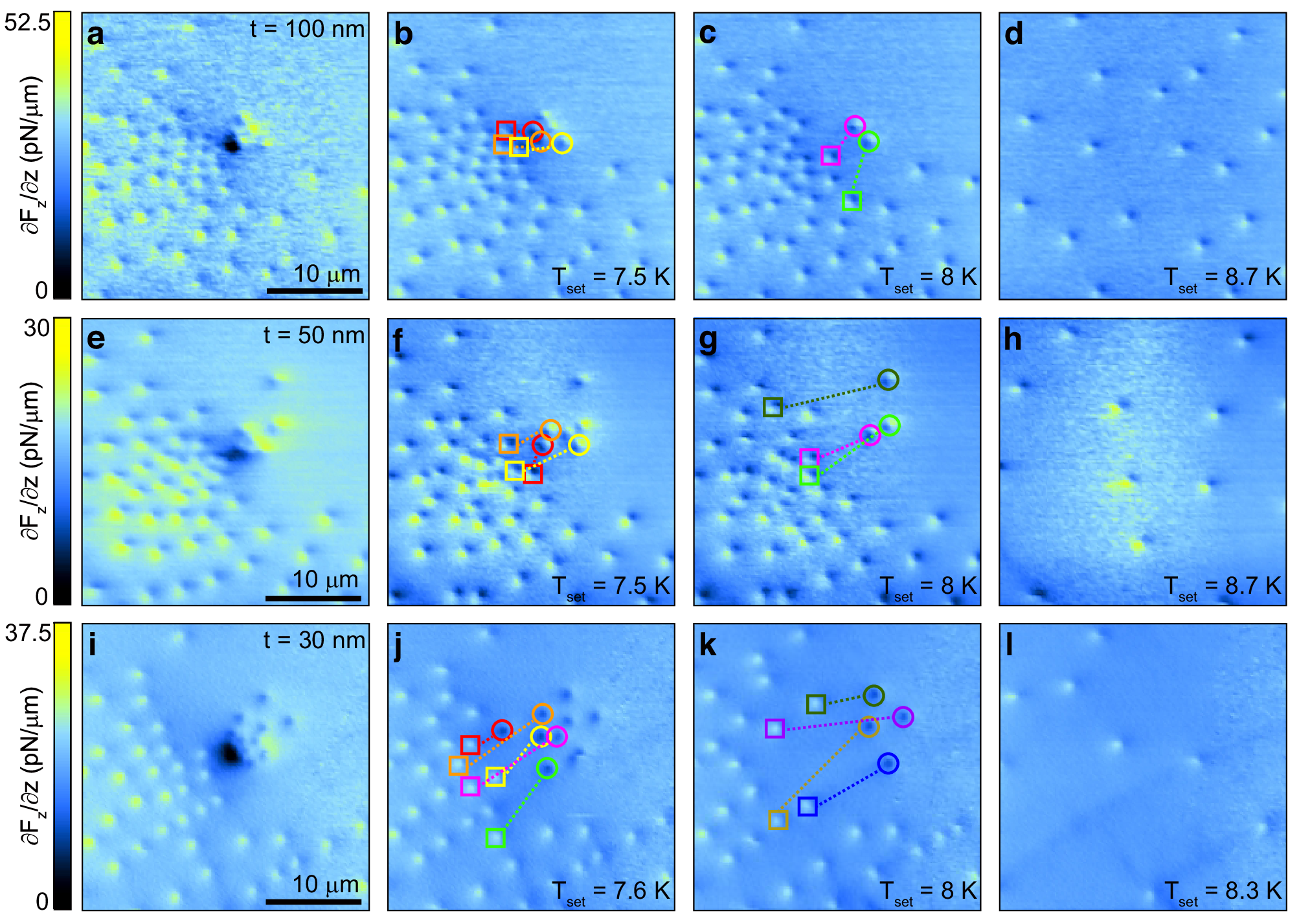}
\caption{Thermal evolution of the SVAPs in thinner Nb films. Thermally assisted depinning and pair annihilation of SVAPs in Nb films of \textbf{a}-\textbf{d} 100 nm, \textbf{e}-\textbf{h} 50 nm, and \textbf{i}-\textbf{l} 30 nm, respectively. Note that the superconducting transition temperatures ($T_{c}$) and magnetic penetration depths ($\lambda$) are different in various Nb films ranged from $\approx$ 8.7 K to 8.3 K and from $\approx$ 90 nm to 145 nm, respectively. The annihilation of each vortex-antivortex pair is traced by the box-circle link with a dotted color line.}
\label{fig5} 
\end{figure*}

In order to investigate the confinement effect of the SVAP connected by an unquantized flux, we examined Nb films with thicknesses smaller than the magnetic penetration depth ($\lambda$). Specifically, we prepared films with thicknesses of 100 nm, 50 nm, and 30 nm, with corresponding $\lambda$ values of 90 nm, 130 nm, and 145 nm, respectively. Among these samples, our focus was on the 30-nm film, which had a $t-\lambda$ ratio of 0.26. This ratio suggests that the 30-nm film does not exhibit a flux quantum along the in-plane direction. Using the same experimental procedure as before, we created SVAPs in these thin films and performed temperature dependence experiments. Surprisingly, we observed pair annihilation not only in the 100-nm and 50-nm films but also in the 30-nm film, as shown in Fig.~\ref{fig5}. The presence of long-range interaction even in the 30-nm film, where $t\ll\lambda$, is intriguing. It indicates that the unquantized magnetic flux\cite{Bardeen1961,Ginzburg1962,Fetter1980,Kogan1994,Geim2000,Zhang2022} associated with the SVAP can mediate long-range forces between superconducting vortices. To substantiate our findings of non-quantized vortices and to closely examine the current distribution within the 30-nm Nb film, we have carried out comprehensive time-dependent Ginzburg-Landau simulations. These simulations have revealed a clear pattern: with decreasing film thickness, there is a proportional reduction in magnetic flux, accompanied by a commensurate increase in current flux. We also examined the contrast differences in MFM images by analyzing the line profiles of each vortex in Nb films with thicknesses of 300 nm ($d$ $>$ $\lambda$) and 30 nm ($d$ $<$ $\lambda$). Unlike the 300-nm film, the 30-nm film shows an intensity difference between isolated and paired vortices, suggesting the formation of unquantized vortices. We refer the supplementary for an in-depth analysis. Such trends affirm a direct relationship between the film's thickness and the manifestation of non-integer quantized flux. This result has significant implications for various applications, including superconducting nanowires, low-dimensional Josephson junctions, and monolayer superconductors.

Moreover, the 1D nature of the SVAP provides a robust platform for exploring the possibilities of braiding operations using unquantized magnetic flux in topological superconducting heterostructure devices. These devices are at the forefront of current research in condensed matter physics and topological quantum computing. The ability to confine and manipulate the unquantized flux within the SVAPs opens up new avenues for studying topological properties and realizing exotic quantum phenomena. Our findings highlight the unique confinement effect of SVAPs connected by unquantized magnetic flux in thin films.

\section*{Conclusion}
Our study demonstrated the creation and manipulation of a vortex-antivortex pair connected by an unquantized magnetic flux in ultra-thin superconducting Nb films using a vector-field MFM. The observed long-range interaction through the unquantized flux provides evidence for the universal formation of such flux regardless of the system's size. This finding contributes to our understanding of superconducting magnetic properties at the nanoscale, where the non-quantization of magnetic flux plays a role in the superconducting properties.

Our results offer an experimental platform for directly investigating a linear potential and its phase transition in a simple 1D fluxoid model system. This has implications for studying the fundamental properties of superconductivity and exploring novel quantum phenomena in low-dimensional systems. Additionally, the long-range interaction mediated by the unquantized magnetic flux opens up possibilities for realizing non-Abelian statistics through the manipulation and annihilation of SVAPs. This has potential implications for the development of topological quantum devices and the field of topological quantum computing.\\

\begin{acknowledgments}
This work was supported by Basic Science Research Program through the National Research Foundation of Korea (NRF) funded by the Ministry of Science, and ICT and Future Planning (NRF-2019R1A2C2090356,NRF-2021R1A6A1A10042944, NRF-2022M319A1073808, and
NRF-2022H1D3A3A01077468). Work at Los Alamos National Laboratory was performed under the auspices of the U.S. Department of Energy. D.W. acknowledges support by the Institute for Basic Science in Korea (Grant No. IBS-R009-Y3). R.M. was funded by US DOE BES program. The authors thank F. Ronning, A. Balatsky, and M.J. Graf for useful discussions and thank the Institute for Basic Science. We appreciate the  APCTP for its hospitality during completion of this work.\\
\end{acknowledgments}

\noindent{\bf Author Contributions}\\
J.K. conceived the original idea and supervised the project. J.Y. G.H. and K.K. performed the analytical and numerical work. G.K., J.Y., I.Y., and J.K. performed the experiments. G.K. made the Nb films. G.K., J.Y., I.Y., D.W., K.K., and J.K. analysed the data, and wrote the manuscript. All authors discussed the results and commented on the manuscript.\\

\noindent{\bf Conflict of interest}\\
The authors declare no competing interests.\\

\clearpage

\clearpage

\pagebreak

\clearpage

\clearpage

\nocite{*}

\begin{thebibliography}{99}
\bibitem{LittlePark1962}
Little, W. A. \& Parks, R. D. Observation of quantum periodicity in the transition temperature of a superconducting cylinder, Phys. Rev. Lett.  $\bf{9}$, 9 (1962).
\bibitem{Harlingen1982}
Van Harlingen,D. J. Thermoelectric effects in the superconducting state, Physica B+C $\bf{109}$, 1710 (1982).
\bibitem{Ginzburg1989}
Ginzburg, V. L. Thermoelectric effects in superconductors, Journal of Superconductivity, $\bf{2}$, 323 (1989).
\bibitem{Ginzburg1991}
Ginzburg,V. L. Thermoelectric effects in superconductors, Supercond. Sci. Technol. $\bf{4}$, S1 (1991).
\bibitem{Josephson1974}
Josephson, B. D. The discovery of tunnelling supercurrents, Rev. Mod. Phys. $\bf{46}$, 251 (1974).
\bibitem{Likharev1979}
Likharev, K. K. Superconducting weak links, Rev. Mod. Phys. $\bf{51}$, 101 (1979).
\bibitem{Jaklevic1964}
Jaklevic, R. C., Lambe, J. Silver, A. H., \& Mercereau,J. E., Quantum interference effects in \uppercase{J}osephson tunneling, Phys. Rev. Lett. $\bf{12}$, 159 (1964).
\bibitem{Kirtley1999}
Kirtley, J. R., Tsuei, C. C. \& Moler, K. A., Temperature dependence of the half-integer magnetic flux quantum, Science $\bf{285}$, 1373 (1999).
\bibitem{Maeno2012}
Maeno,Y., Kittaka, S., Nomura, T. Yonezawa, S. \& Ishida, K. Evaluation of spin-triplet superconductivity in \uppercase{S}r$_{2}$\uppercase{R}u\uppercase{O}$_{4}$, J. Phys. Soc. Jpn. $\bf{81}$, 011009 (2012).
\bibitem{Kawabata2007}
Kawabata, S., Kashiwaya, S., Asano, Y., Tanaka, Y., Kato, T., and A Golubov, A., Supercond. Sci. Technol. $\bf{20}$, S6 (2007).
\bibitem{Brandt1993}
Brandt, E. H. Indenbom, M. V.  and Forkl, A., Type-\uppercase\expandafter{\romannumeral2} superconducting strip in perpendicular magnetic field, EPL $\bf{22}$, 735 (1993).
\bibitem{Stenuit2004}
Stenuit, G., Michotte, S., Govaerts, J. \& Piraux, L. Temperature dependence of penetration and coherence lengths in lead nanowires, Supercond. Sci. Technol.  $\bf{18}$, 174 (2005).
\bibitem{Engbarth2011}
Engbarth, M. A., Bending, S. J. \& Milošević, M. V. Geometry-driven vortex states in type-\uppercase\expandafter{\romannumeral1} superconducting \uppercase{P}b nanowires, Phys. Rev. B $\bf{83}$, 224504 (2011).
\bibitem{Geim2000}
Geim, A. K., Dubonos, S. V., Grigorieva, I. V., Novoselov, K. S., Peeters, F. M., and Schweigert, V. A., Non-quantized penetration of magnetic field in the vortex state of superconductors, Nature $\bf{407}$, 55 (2000).
\bibitem{Zhang2022}
Zhang, A-L., Gladilin, V., Van de Vondel, J., Moshchalkov, V. V. \& Ge, J-Y. Tunable noninteger flux quantum of vortices in superconducting strips, Nano. Lett. $\bf{22}$, 7151-7157 (2022).
\bibitem{Bardeen1961}
Bardeen, J. Quantization of flux in a superconducting cylinder, Phys. Rev. Lett. $\bf{7}$, 162 (1961).
\bibitem{Ginzburg1962}
Ginzburg, V. L. Magnetic flux quantization in a superconducting cylinder, Sov. Phys. JETP. $\bf{15}$, 207 (1962).
\bibitem{Fetter1980}
Fetter, A. L. Flux penetration in a thin superconducting disk, Phys. Rev. B $\bf{22}$, 1200 (1980).
\bibitem{Kogan1994}
Kogan, V. G. Pearl’s vortex near the film edge, Phys. Rev. B. $\bf{49}$, 15874 (1994).
\bibitem{yang2016construction}
Yang, J., Yang, I., Kim, Y. W., Dongwoo Shin, D., Juyoung Jeong, J., Wulferding, D., et al. Construction of a $^{3}$\uppercase{H}e magnetic force microscope with a vector magnet, Rev Sci Instrum $\bf{87}$, 023704 (2016).
\bibitem{nazaretski2009direct}
Nazaretski, E., Thibodaux, J. P., Vekhter, I., Civale, L., Thompson, J. D., Movshovich, R. Direct measurements of the penetration depth in a superconducting film using magnetic force microscopy,Appl. Phys. Lett $\bf{95}$, 262502 (2009).
\bibitem{Kimlambda2012}
Kim, J., Civale, L., Nazaretski, E., Haberkorn, N., Ronning, F., Sefat, A. S., et al. Direct measurement of the magnetic penetration depth by magnetic force microscopy, Supercond. Sci. Technol. $\bf{25}$, 112001 (2012).
\bibitem{Blatter1994}
Blatter, G., Feigel’man, M. V., Geshkenbein, V. B., Larkin, A. I. \& Vinokur, V. M. Vortices in high-temperature superconductors, Rev. Mod. Phys. $\bf{66}$, 1125 (1994).
\bibitem{tinkham2004introduction}
Tinkham, M. Introduction to superconductivity (Courier Corporation, 2004).
\bibitem{Auslaender2009}
Auslaender, O. M., Luan, L., Straver, E., Hoffman, J. E., Koshnick, N. C., Zeldov, E., et al. Mechanics of individual isolated vortices in a cuprate superconductor, Nat. Phys. $\bf{5}$, 35-39
(2009).
\bibitem{Kremen2016}
Kremen, A., Wissberg, S., Haham, N., Persky, E., Frenkel, Y., and Kalisky, B. Mechanical control of individual superconducting vortices, Nano. Lett. $\bf{16}$, 1626-1630
(2016).
\bibitem{Ge2016}
Ge, J-Y., Gladilin, V. N., Tempere, J., Xue, C., Devreese, J. T., De Vondel, J. V., Zhou, Y., et al. Nanoscale assembly of superconducting vortices with scanning tunneling microscope tip, Nat. Commun. $\bf{7}$, 13880
(2016).
\bibitem{Wulferding2020}
Wulferding, D., Kim, G., Kim, H., Yang, I., Bauer, E. D., Ronning, F., et al. Local characterization of a heavy-fermion superconductor via sub-\uppercase{K}elvin magnetic force microscopy, Appl. Phys. Lett. $\bf{117}$, 252601 (2020).

\end{thebibliography}
\end{document}